\documentstyle[prd,tighten,floats,preprint,aps]{revtex}
\begin{document}
\draft
\preprint{UPR-721-T}
\date{October 1996}
\title{Constant threshold correction to electrically charged dilatonic black holes}
\author{Kwan-Leung Chan
\thanks{E-mail address: klchan@cvetic.hep.upenn.edu}}
\address{Department of Physics and Astronomy \\
         University of Pennsylvania, Philadelphia PA 19104-6396}
\maketitle
\begin{abstract}
{We investigate the effect of a constant threshold correction to a general non-extreme, static, spherically symmetric, electrically charged black hole solution of the dilatonic Einstein-Maxwell Lagrangian, with an arbitrary coupling $\beta$ between the electromagnetic tensor and the dilaton field. For a small $\beta$, an exact analytical solution is obtained. For an arbitrary $\beta$, a close form solution, up to first order in the constant threshold correction, of the metric and the dilaton are presented. In the extremal limit, the close form solution is reduced to an exact analytical form.}
\end{abstract}
\vskip 1.5cm

Dilaton, i.e., scalar field without self-interactions, arise naturally in basic theories that unify gravity with other interactions, including certain supergravity theories, and effective theory from superstrings. In general, it couples to the gauge field kinetic energy as well as to the matter potential. It is therefore of interest to address dilatonic topological configurations in such theories. The dilatonic charged black hole solutions without threshold correction are known \cite{Gary} - \cite{Dark}. The presence of the coupling between dilaton and the electromagnetic tensor produces black hole solutions drastically different from the Reissner-Nordstrom solutions. 
\newline

In addition to the dilaton, there are moduli fields which are generically present in string theories. They are stringy modes in a vacuum associated with compactification of the extra dimensions. In general, they act as threshold corrections to the scalar function that couples to the gauge field kinetic energy \cite{CF} \cite{Lust}. Such scalar function determines the strength of the effective gauge coupling constant. 
\newline

Effects of stringy threshold corrections on charged spherically symmetric dilatonic configurations without gravity have been studied in \cite{CT}. The present work is to generalize the study to charged black hole configurations, i.e., by including gravitational effects. 
\newline

The most general bosonic Lagrangian for the dilatonic Einstein-Maxwell theory with just one moduli field is of the following form:

\begin{equation}
{\cal L} = \int d^4 x {\sqrt -g} \left[ -R + 2 \partial_{\mu}\phi \partial ^{\mu}\phi + 2 \partial_{\mu}\varphi \partial^{\mu}\varphi + f( \phi , \varphi ) F^2  - V( \phi , \varphi ) \right] ,
\label{GENERAL}
\end{equation}
where $F_{\mu \nu}$ is the electromagentic tensor field strength, $\phi, \varphi$ are the dilaton field and the moduli field, respectively. The potential $V( \phi, \varphi)$ is expected to be non-perturbative as both the dilaton and the moduli field are flat directions of string theories when studied in terms of perturbation theory. 
\newline

As a first step, we assume that the threshold correction can be approximated by a small constant, c. Therefore we neglect the kinetic term of $\varphi$, and take the gauge coupling function as

\begin{equation}
f( \phi , \varphi ) = e^{-2 \beta \phi} + c ,
\label{GCF}
\end{equation}
where $\beta$ is an arbitrary parameter
\footnote{The case with $\beta = 1$ and a running moduli field was studied in \cite{Mignemi}.}. 
In the compactified supergravity models associated with the low-energy limit of superstring theories, there are several different consistent truncations which give the dilatonic Einstein-Maxwell Lagrangians (of the form (\ref{GENERAL}), but without $V, c$, and $\varphi$) with $\beta = 0, {1 \over \sqrt{3}}, 1$, and $\sqrt{3}$ \cite{Duff1} \cite{Duff2}. Here we work with an arbitrary $\beta$, and so include more general supergravity theories. Furthermore, we will neglect the potential, $V$, and the terms with higher order derivatives. We assume that the size of our black hole is much bigger than Planck length
\footnote{Higher order corrections, ${\it i.e.}$, $\alpha'$ correction, is considered in \cite{Makoto}.}. 
\newline

Therefore we work on the Lagrangian 

\begin{equation}
{\cal L} = \int d^4 x {\sqrt -g} \left[ -R + 2 \partial_{\mu}\phi \partial ^{\mu}\phi + f( \phi )  F^2  \right] ,
\label{START}
\end{equation}
with $f$ taken from (\ref{GCF}).
\newline

We will study the electrically charged solution only, as we would like to work on situations in which the threshold corrections can always be treated perturbatively. For the electrically charged solution, $\phi (r) \rightarrow - \infty$ as $r \rightarrow \alpha$, where $\alpha$ is the (inner) horizon, when threshold correction is totally neglected. So we expect that $f$ is dominated by $e^{-2 \beta \phi}$ for all values of r even when the small threshold correction is taken into account, and we can approximate the exact solution with a perturbation series in c 
\footnote{We set the asymtotic value of dilaton to unity to simplify the formulae}. 
On the contrary, $\phi (r) \rightarrow \infty$ as $r \rightarrow \alpha$ in an magnetically charged solution when threshold correction is neglected, so $f$ is expected to be dominated by the threshold correction as $r$ approaches to the horizon when c is non-zero, and perturbation in c would not be justified. 
\newline

It is more convenient to work on a gauge couping function which is normalized to unity as $r \rightarrow \infty$. Therefore instead of taking $f$ from (\ref{GCF}), we work on $f = f_N$, with 

\begin{equation}
f_N ( \phi ) \equiv { {e^{-2 \beta \phi} + c} \over {1 + c} } .
\label{NORF}
\end{equation}
We will first solve the Euler-Lagrange equations from the Lagrangian (\ref{START}), but with the gauge coupling function, $f = f_N$ defined in (\ref{NORF}). Then, we identify the charge of the electromagnetic field, $Q_c$, as:

\begin{equation}
Q_c = Q_o \sqrt{1+c} ,
\label{QC}
\end{equation}
where $Q_o$ is the charge when $c=0$, to get the solution of the Lagrangian (\ref{START}) with the gauge coupling function, $f$ given by (\ref{GCF}).  
\newline

We take static, spherically symmetric ansart for the metric:
\begin{equation}
ds^2 = - {\lambda}^2(r) dt^2 + {\lambda}^{-2}(r) dr^2 + R^2(r) d{\Omega} .
\label{METRIC}
\end{equation}
The dilaton depends on $r$ only from spherical symmetry. The electromagnetic tensor for an electric solution is:
\begin{equation}
F = {Q_c \over {R^2 f_N ( \phi ) } }  dt \ \ \bigwedge \ \ dr .
\label{TENSORN}
\end{equation}
As $f_N$ is normalized, and we expect $R^2 \rightarrow r^2$ asymtotically, so $Q_c$ is the physical electric charge of the solution. That is the advantage of taking $f = f_N$, intead of taking $f$ from (\ref{GCF}) directly. The equations of motion are

\begin{equation}
{\lambda}^2 R^2 = r^2 ( 1 - a x ) ( 1 - x ) ,
\label{LAMXR}
\end{equation}

\begin{equation}
( 1 - x ) ( 1 - a x ) \left[ ( 1 - x ) ( 1 - a x ) \varphi ' \right] ' = 2 ( 1 + c ) { Q_c^2 \over \alpha^2 }{ {Z  e^{ (1 + \beta^2) \varphi }} \over ( 1 + c e^{\beta^2 \varphi} )^2 } ,
\label{MASONE}
\end{equation}

\begin{equation}
\left[ ( 1 - x ) ( 1 - a x ) { Z' \over Z } \right]' = c e^{\beta^2 \varphi} \left[ ( 1 - x ) ( 1 - a x ) \varphi ' \right]' ,
\label{MASTWO}
\end{equation}
where
\begin{equation}
\varphi \equiv { {2 \phi} \over \beta } ,
\label{VARPHI}
\end{equation}

\begin{equation}
{Z} = {\lambda}^2 {\rm e}^{-{ \varphi }} ,
\label{DEFZ}
\end{equation}
and $ x \equiv {\alpha \over r} $, $ a \equiv {r_+ \over r_-} $, $ \alpha \equiv r_- $, where $r_-, r_+$ are respectively the inner and outer horizon. Note that $\varphi$ in the above equations are not the moduli fields mentioned at the beginning of this paper. Here $\varphi$ is defined in (\ref{VARPHI}).
\newline

For small $\beta$, {\it i.e.}, when $\beta^2$ is ignored, the equations (\ref{MASONE}) and (\ref{MASTWO}) are exactly solvable.

\begin{equation}
\lambda^2 = ( 1 - x ) ( 1 - a x ) ,
\label{RNMETRIC}
\end{equation}

\begin{equation}
\phi = { \beta \over { 1 + c } } \ln ( 1 - x ) ,
\label{RNPHI}
\end{equation}

\begin{equation}
\alpha^2 = { Q_c^2 \over a } ,
\label{RNALPHA}
\end{equation}
With $Q_c$ given by (\ref{QC}), the above is the solution of the Lagrangian (\ref{START}) with the gauge coupling function $f$ given by (\ref{GCF}). The mass of the black hole is: $M = { {1 + a} \over {2 \sqrt{a}} } Q_c$. The inner horizon $r_-$ is given by: $r_- = \alpha = { Q_c \over \sqrt{a} }$, and the outer horizon ,$r_+$, is given by: $r_+ = a r_-$. 
\newline

Therefore the mass of the black hole increases and the horizons are pushed outward when threshold correction is taken into account. In the limit as $\beta \rightarrow 0$ with arbitrary c, we recover the expected Reissner-Nordstrom solution with a vanishing dilaton, as required by the no-hair theorem. It should be noted that as long as $\beta$ is non-zero, the singularity is still at the inner horizon (${\it i.e.}, r = r_-$). The dilaton diverges as $x \rightarrow 1$ no matter how big $c$ is, in spite of the fact that the metric $\lambda$ has the same form as the Reissner-Nordstrom metric. That is because we have neglected the terms with the order of $\beta^2$. 
\newline

We now consider the case when $\beta^2$ can be arbitrary. We use first order perturbation theory in $c$ to study the change of the metric and the dilaton by the constant threshold correction. 
\newline

We expand the metric $Z$ and the scaled dilaton $ \varphi $ around $ c = 0 $ up to 1st order

\begin{equation}
\varphi = \varphi_o + c \varphi_1 ,
\label{PERVP}
\end{equation}

\begin{equation}
Z = Z_o + c Z_1 ,
\label{PERZ}
\end{equation}

\begin{equation}
\lambda^2 = \lambda^2_o + c \lambda^2_1 ,
\label{PERLAM}
\end{equation}

From the zeroth order equations of (\ref{MASONE}) and (\ref{MASTWO}), and use (\ref{DEFZ}) to relate $Z_o$ to $\lambda^2_o$, we get the expected GHS solution

\begin{equation}
{\lambda}^{2}_0 = ( 1 - a x ) ( 1 - x )^{{1 - \beta ^2}\over {1 + \beta ^2}} ,
\label{GHSLAM}
\end{equation}

\begin{equation}
{\rm e}^{2 \phi_0} = ( 1 - x )^{{2 \beta}\over {1 + \beta ^2}} ,
\label{GHSPHI}
\end{equation}

\begin{equation}
\alpha ^2 = {1 \over a} ( 1 + \beta ^2) Q^2 ,
\label{GHSAL}
\end{equation}

\begin{equation}
Q = Q_o .
\label{GHSQ}
\end{equation}

Note that as we work on the first order corrections for the metric and the dilaton in the following, we do not consider the dependence on $c$ of the electric charge, $Q_c$, as explained in previous paragraphs. After we have gotten the solutions, we "reinsert" the c-dependence of $Q_c$ from (\ref{QC}). From (\ref{MASTWO}), the first order correction of $Z$ is:

\begin{equation}
{Z}_1(x) = {{K_1(1-ax)}\over {(1-a)(1-x)}} \ln {{1-ax}\over{1-x}} - {{2a(1-ax)}\over{(1+3 \beta^2)(1-x)}} \int^x_0 dy \left[{ {(1-y)^{{2\beta^2}\over{1+\beta^2}}} \over {1-ay}} \right] ,
\label{ZONE}
\end{equation}
One of the two integration constants expected from the 2nd order linear differential equation (\ref{MASTWO}) has been fixed by requiring that $Z_1 \rightarrow 0$ as $x \rightarrow 0$. Such boundary condition is equivalent to require both $\lambda^2_1 \rightarrow 0$ and $\varphi \rightarrow 0$ asymtotically. In fact, from (\ref{DEFZ}), and (\ref{PERZ}), (\ref{PERLAM}), we find

\begin{equation}
{Z}_1 = {\rm e}^{- \varphi_o} \left( \lambda^2_1 - \varphi_1 \lambda^2_0 \right) .
\label{Z1LAM1}
\end{equation}

To fix $K_1$, we require that as $x \rightarrow {1 \over a}$, $\lambda^2_1 \over \lambda^2_o$ converges (though both $\lambda_o$ and $\lambda_1$ vanish at the horizons), which should be reasonable for perturbation theory to be applicable. So we get

\begin{equation}
K_1 = {{2(a-1)}\over {1+3\beta^2}} \left({ {a-1}\over a} \right)^{{2\beta^2}\over{1+\beta^2}} .
\label{K1FIXED}
\end{equation}

The first order equation from (\ref{MASONE}) is

\begin{equation}
{\varphi}''_1 + {{2ax-1-a}\over {(1-x) (1-ax)}} \varphi'_1 - {{2a}\over {(1-x) (1-ax)}} \varphi_1 = R(x) ,
\label{VAR2ND}
\end{equation}
where

\begin{equation}
R(x) = {2 a \over {(1+\beta^2) (1-ax)^2}} \left[ {\cal Z}_1 - 2 c (1-ax) (1-x)^{{\beta^2-1}\over {\beta^2+1}}  + { {1-a x} \over {1-x}} \right] .
\label{SOURCE}
\end{equation}
This equation has close form solution 
\begin{equation}
\varphi_1 (x) = K_3 \varphi^h_1 (x) + K_4 \varphi^h_2 (x) + \varphi_p (x) ,
\label{VARMAS}
\end{equation}
where $\varphi^h$ are solutions of the homogeneous equation, given by
\begin{equation}
\varphi^h_1(x) = {1 \over 2} ( 1 + a - 2 a x ) ,
\end{equation}

\begin{equation}
\varphi^h_2(x) = {6 \over(1-a)^3}\left[2(a-1)+(1+a-2ax)\ln{{1-ax}\over{a(1-x)}} \right] .
\end{equation}
They have been normalized so that as $a \rightarrow 1$, they reduce to the forms: $ 1-x $ and $ 1/(1-x)^2 $. The particular solution is

\begin{equation}
\varphi_p = \varphi^o_p - {1 \over {1 + \beta^2} } ,
\label{VARPMAS}
\end{equation}
where

\begin{equation}
\varphi_p^o (x) = {{2a} \over {3(1+\beta^2)}} \left[ - \varphi^h_1(x) \int ^x_0 dy \left( h(y) \varphi^h_2 (y) \right) + \varphi^h_2 (x) \int^x_0 dy \left( h(y) \varphi^h_1(y) \right) \right] ,
\label{VARO}
\end{equation}

\begin{equation}
h(x) = {{1-x}\over {1-ax}} \left[ {\cal Z}_1 (x) - 2c(1-ax)(1-x)^{{\beta^2-1}\over {\beta^2+1}} \right] .
\end{equation}

Requiring $\varphi( x \rightarrow 0 ) \rightarrow 0$ relate $K_3$ and $K_4$ as

\begin{equation}
K_3 = K_4 {{12}\over{(1-a)^3 (1+a)}} \left[ (1+a)\ln a - 2(a-1) \right] + { 2 \over { ( 1 + a ) ( 1 + \beta^2 )} } .
\label{K3K4}
\end{equation}
As $x = { 1 \over a }$ is equivalent to $ r = a \alpha $ which is a horizon, but not a singularity, we expect $\varphi_1$, like $\varphi_o$, does not diverge there. This fixes $K_4$ as

\begin{equation}
K_4 = {-a\over {3(1+\beta^2)}} \int ^{1\over a}_0 dx ( 1 + a - 2 a x ) \left[ { {1 - y} \over {1 - a y} } Z_1(y) - 2 ( 1 - y )^{ {2 \beta^2} \over {1 + \beta^2}} \right] .
\label{K4CLOSE}
\end{equation}
So we have obtained the first order correction in c of the dilaton from (\ref{VARMAS}) to (\ref{VARO}), while the first order correction to the metric is from (\ref{ZONE}) to (\ref{K1FIXED}). The inner horizon, $\alpha$, is still given by (\ref{GHSAL}), but with $Q = Q_c$, instead of $Q = Q_o$ in (\ref{GHSQ}). The outer horizon is given by: $r_+ = a r_-$.
\newline

In the extremal limit, {\it i.e.}, $a \rightarrow 1$, the above equations for $Z_1, \varphi_1$ have exact analytical forms as follows:

\begin{equation}
\varphi_1 = { {( 1 + \beta^2) (1 - 3 \beta^2)} \over { \beta^2 (1 - \beta^2) (1 + 3 \beta^2)} } \left[ 1 - ( 1 - x )^{ {2 \beta^2} \over {1 + \beta^2}} \right] + \left( { {2 \beta^2} \over { (1+\beta^2) (1-\beta^2)} } \right) x ,
\label{VAR1A1}
\end{equation}

\begin{equation}
Z_1 = { {1+\beta^2} \over { \beta^2 (1+3 \beta^2)} } \left[ (1-x)^{ {2 \beta^2} \over {1+\beta^2} } - 1 \right] .
\label{Z1A1}
\end{equation}
They imply

\begin{equation}
\lambda^2_1 = (1-x)^{2 \over {1+\beta^2}} \left[ { {2 \beta^2} \over { (1+\beta^2)(1-\beta^2) } } x - { {2(1+\beta^2)} \over {(1+ 3 \beta^2)(1-\beta^2)}} \left( 1-(1-x)^{ {2 \beta^2} \over {1+\beta^2}} \right) \right] .
\label{LAM1A1}
\end{equation}

In the limit of : $\beta^2 \rightarrow 0$, the first order dilaton and metric from (\ref{VAR1A1}), (\ref{Z1A1}), and (\ref{LAM1A1}) agree with that obtained by expanding the metric and dilaton from (\ref{RNMETRIC}) and (\ref{RNPHI}) with $a \rightarrow 1$. 
\newline

With the analytical forms (\ref{VAR1A1}) and (\ref{LAM1A1}), we find that the mass, $M$, of the black hole and the charge of the dilaton, $D$, are:

\begin{equation}
M = \alpha \left[ {1 \over { 1+\beta^2 }} + c { {\beta^2} \over { (1+3 \beta^2)(1+\beta^2) }} \right] ,
\label{MA1}
\end{equation}

\begin{equation}
D = \alpha \left[ {\beta \over { (1+\beta^2) }} - c { { \beta} \over { (1+3 \beta^2)(1+\beta^2) }} \right] ,
\label{DA1}
\end{equation}
respectively. Recall that the physical electric charge is given by $Q_c$ from (\ref{QC}), and $\alpha$ is given by (\ref{GHSAL}) with $Q \equiv Q_c$. Up to first order in $c$, we find: $ Q_1 Q_2 - M_1 M_2 - D_1 D_2 = 0 $, where $Q_i$, $M_i$, and $D_i$ are the physical electric charge, mass, and dilatonic charge of an extremal electrically charged black hole labelled by $i$, respectively. Therefore the repulsive force between any two black holes exactly balances the attractive forces from gravity and that produced by the dilaton fields. Multi-black hole solutions are thus possible in the extremal limit, just like the case with no threshold correction \cite{Dark}. 
\newline

Therefore, the dilatonic extremal black hole perturbed by a small constant threshold correction has its mass increased and its horizon pushed outward. Like the case without the threshold correction, the singularity surface coincide with the horizon, and so the black hole has zero entropy.

\section{Acknowledgments}

I am grateful to Mirjam Cveti\v c for many useful and enlightening discussions. The work was supported in part by U.S. Department of Energy Grant No. DOE-EY-76-02-3071 and the National Science Foundation PHY95-12732.

\end{document}